# Sensitivity analysis of biological washout and depth selection for a machine learning-based dose verification framework in proton therapy

Shixiong Yu[1], Yuxiang Liu[1], Zongsheng Hu[2], Haozhao Zhang[3], Pengyu Qi[4*], Hao Peng[3*]

[1] Department of Medical Physics, Wuhan University, Wuhan, China, 430072

[2] Department of Radiation Physics, University of Texas MD Anderson Cancer Center, Houston, TX, USA, 77573

[3] Medical Artificial Intelligence and Automation (MAIA) Laboratory, Department of Radiation Oncology, University of Texas Southwestern Medical Center, Dallas, TX, USA, 75390

[4] Institute of Biomedical Engineering, Shenzhen Bay Laboratory, Shenzhen, China, 518107

**\*Corresponding Authors:**

Qiyu Peng, PhD., Pengqy@szbl.ac.cn

Hao Peng, PhD., Hao.peng@utsouthwestern.edu

**Running title:**

Biological washout and depth selection for online verification

**Conflict of interest:**

None

**Acknowledgement:**

None

**Objective:** Dose verification based on proton-induced positron emitters is a promising quality assurance tool and may leverage the strength of artificial intelligence. To move a step closer towards practical application, the sensitivity analysis of two factors needs to be performed: biological washout and depth selection.

**Approach:** A bi-directional recurrent neural network (RNN) model was developed. The training dataset was generated based upon a CT image-based phantom (abdomen region) and multiple beam energies/pathways, using Monte-Carlo simulation (1 mm spatial resolution, no biological washout). For the modeling of biological washout, a simplified analytical model was applied to change raw activity profiles over a period of 5 minutes, incorporating both physical decay and biological washout. For the study of depth selection (a challenge linked to multi-field/angle irradiation), truncations were applied at different window lengths (100, 125, 150 mm) to raw activity profiles. Finally, the performance of a worst-case scenario was examined by combining both factors (depth selection: 125 mm, biological washout: 5 mins). The accuracy was quantitatively evaluated in terms of $\Delta_{range}$, mean absolute error (MAE) and mean relative errors (MRE).

**Main results:** The machine learning model is found to be more sensitive to depth selection, relative to biological washout. For the worst-case, the mean $\Delta_{range}$ and MRE95 are 6.26 mm and 10.1% when only activity is used as input, respectively. After including anatomical and stopping power prior as auxiliary inputs, the verification accuracy is significantly improved ($\Delta_{range}$ : 0.38 mm, MRE95: 3.11%).

**Significance:** Our proposed AI framework shows good immunity to the perturbation associated with two factors. The detection of proton-induced positron emitters, combined with machine learning, has great potential to implement online patient-specific verification in proton therapy.

# 1. Introduction

Online dose verification in proton therapy relies on the correlation between the spatial distribution of proton-induced secondary signals and dose deposition. Secondary signals stay within three categories: positron emitters (i.e. positron emission tomography, PET) (Parodi et al 2002, Nishio et al 2005, Lu et al 2008, Fiedler et al 2010, Espana et al 2010, Espana et al 2011, Helmbrecht et al 2012), prompt gamma (Bom et al 2012, Gueth et al 2013) and acoustic signa (Jones et al 2018, Yu et al 2019, Yao et al 2020, Yu et al 2021). At the heart of verification is the comparison between a reference established before treatment and a measurement during treatment. The reference can be either obtained from either previous treatment (Nishio et al 2010), or from Monte Carlo simulation (Parodi et al 2007, Parodi et al 2008, Seravalli et al 2012, Robert et al 2013).

Machine learning-based approaches have recently been proposed for online dose verification in proton therapy. For instance, a recurrent neural network (RNN) is able to exploit sequential connections to keep track of the context in an input-output pair, well suited for the purpose of dose verification (Li et al 2019, Liu et al 2019, Hu et al 2020). When treating verification as a mapping between activity and dose, a generative adversarial network (GAN) has also proven its value (Ma et al 2020). Beside the information of proton-induced positron emitters, the inclusion of anatomical information as auxiliary features (e.g. HU from CT images, pre-calculated stopping power), is a valid strategy to ease the workload of machine learning (Hu et al 2020, Ma et al 2020).

Conceivably, when inaccurate activity-dose pairs are provided for training the machine learning model, the prediction will also be comprised. Therefore, the ultimate accuracy of online verification relies on two aspects: 1) the generation of high-fidelity activity-dose pairs for training, and 2) additional factors influencing activity profiles such as biological washout and PET image reconstruction (as well as motion). This manuscript partially addressed the second aspect. The sensitivity analysis targeted for the first aspect has previously been conducted (Knopf et al 2009, Espana et al 2010, Espana et al 2011). For instance, different cross section values of reaction channels were studied with phantom measurement, and the range difference is found to be ~ 1 mm (a 5-min PET scan) and ~ 5 mm (for a 30 min PET scan), with 15 min delay between irradiation and scan (Espana et al 2011). Similarly, HU-based tissue composition and different widths of HU bins were studied in head and neck patients for sensitivity analysis (Espana et al 2010), showing up to ~1 mm range uncertainty and <5% amplitude change in activity profiles. How ~5% fluctuation in activity profiles translates into the ultimate accuracy of verification remains an intriguing question to be answered.

Built upon our previous studies, this paper intends to conduct a sensitivity analysis of two factors which potentially impact the performance of machine learning: biological washout and depth selection. In essence, we aim to understand how sensitive a machine learning model is to any fluctuation/perturbation caused by the two factors, both of which alter the shape of activity profiles and thus confound the underlying correlation between activity and dose profiles. To elaborate, the influence of depth selection is relevant in a situation when there is overlap of activity distribution over a planning treatment volume (PTV), due to multiple irradiation fields (typically 2-4 fields in clinical proton therapy). When proton beams sequentially irradiate a target from two fields, two activity distributions (one for each field) will smear together over the PTV. It is of great intertest to find out whether one can use only a sub-region of activity profiles (i.e. integral depth dose) as input for AI-based verification. If that is indeed the case, the verification can then be simply performed field by field. Said differently, we intend to test the following assumption: a proximal activity distribution holds sufficient information to allow the AI model to predict a complete dose profile (both proximal and distal ends).

It is well known that modeling the decay behaviors of proton-induced positron emitters (mostly $^{15}$O and $^{11}$C) is a challenging task, comprising two processes: physical decay and biological washout. The former one depends only on the types of radionuclides and is straightforward, as done in time-activity correction in standard PET imaging. The latter one is more complicated, since the proton-induced positron emitters are not static but will immediately diffuse/perfuse in living tissues. More important, differing from tracers used in standard PET imaging such as FDG, how positron emitters bind to different molecules and undergo different physiological pathways remains unclear in proton therapy (Mizuno et al 2003, Hirano et al 2013, Ammar et al 2014). By viewing the signal change caused by biological washout as a form of external perturbation, we attempt to investigate to what extent it impacts the performance of the machine learning model.

The sensitivity analysis conducted in this study is expected to shed valuable light on numerous aspects related to implementation of AI-based online verification. For brevity, how the RNN model works for the purpose of verification (e.g. activity+HU+SP in particular) was not discussed in details in this manuscript, but can be found in our previous publications (Liu et al 2019, Hu et al 2020, Ma et al 2020). To avoid potential confusion, the following three issues also need to be clarified. First, our approach is devised to perform signal acquisition for 5 minutes in order to acquire sufficient counts and as such, it may be considered as a semi-online or in-room verification scenario (Zhu et al 2011). Strictly speaking, a truly online scenario performs measurement only during the period of beam delivery. Second, our framework verifies both range and dose, since it can obtain not only the position of a distal fall-off, but also dose information over different depths. Third, as a standard data pre-processing technique in machine learning, all input and output values need to be normalized (e.g. divided by a global normalization factor) to make them stay in the range between 0 and 1, easing the workload of machine learning. However, the AI model is indeed able to provide the information of absolute dose in the unit of Gray, by multiplying the relative dose outputs with the normalization factor.

## 2. Methods and Materials

**2.1 Data Generation**

The modeling of $^{11}$C and $^{15}$O production was performed using GATE (Jan et al 2011). For each mono-energetic beamlet, the energy spectrum was assumed to have a Gaussian distribution with a standard deviation of 1%. The beam profile was assumed to be Gaussian distribution (lateral, sigma: 5 mm) at the nozzle. The distance between the origin of the beam and the center of the target was 0.5 m. The physics list used was QGSP-BIC-HP-EMY. The Urban multiple scattering model and high-precision neutron model were included. The "Dose Actor" tool in GATE was used to record dose deposition, and the "Production and Stopping Actor" tool in GATE was used to record distribution of positron emitters ($^{11}$C and $^{15}$O). Neither physical decay nor biological washout was modeled in Monte-Carlo simulation. For each sample in the dataset, multiple mono-energetic pencil beams were combined to form a spread-out Bragg peak (SOBP), with plateau widths in the range between 15 mm and 55 mm.

The training dataset was generated based upon a CT image-based phantom (abdomen region) and various beam energies, as shown in Fig. 1A. The voxel size was $1 \times 1 \times 1$ mm$^3$. The total number of protons was fixed at $0.5 \times 10^9$ for each sample, corresponding to ~1-2 Gy. The total acquisition window was assumed to be 5 mins for two reasons. First, the period of 5 mins is long enough relative to the half-live values ($^{15}$O: 122.44 seconds and $^{11}$C: 20.38 minutes) (Fig. 1B). Second, a verification routine taking up longer than 5 mins comprises the utilization of treatment room and reduces patient throughput.

The simulated activity profiles refer to the raw distribution of positron emitters directly from Monte-Carlo simulation (1 mm spatial resolution). Neither the interaction of 511 keV gamma rays with tissues nor their detection was considered, which has already been evaluated in our previous work (Hu et al 2020). For simplification, limited spatial resolution was modeled analytically by smoothing the raw activity profiles with a Gaussian kernel (full-width-half-maximum: 4 mm, spatially invariant).

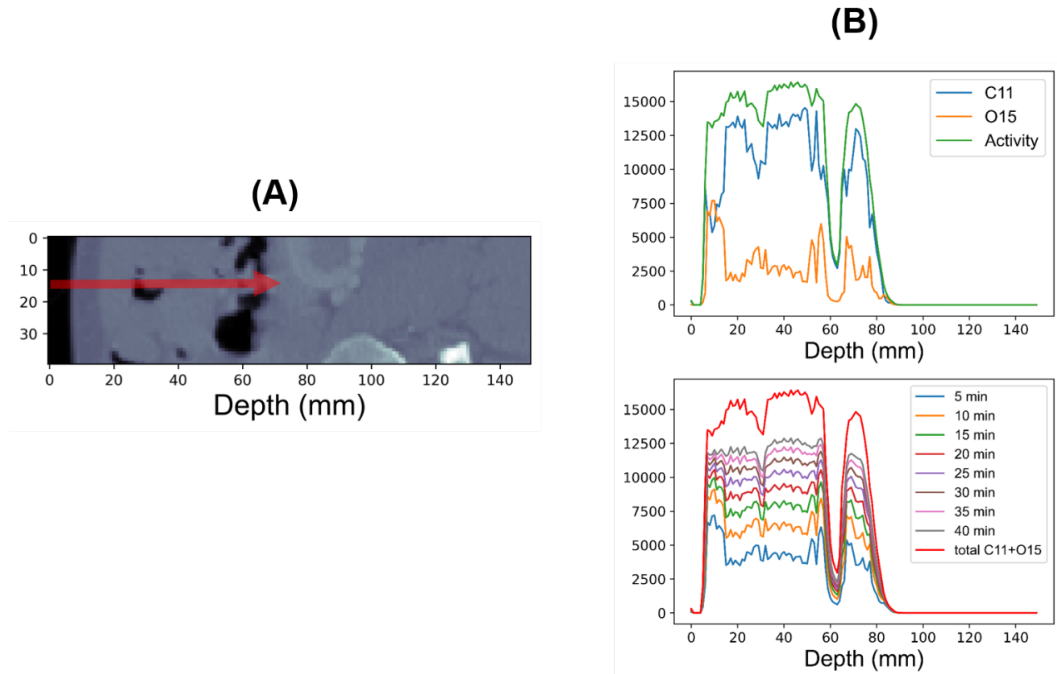

**Fig 1. (A)** Illustration of beam irradiation on the abdomen site. **(B)** The counts of positron emitters ($^{15}$O and $^{11}$C) across the central axis of the beamlet, as a function of acquisition time.

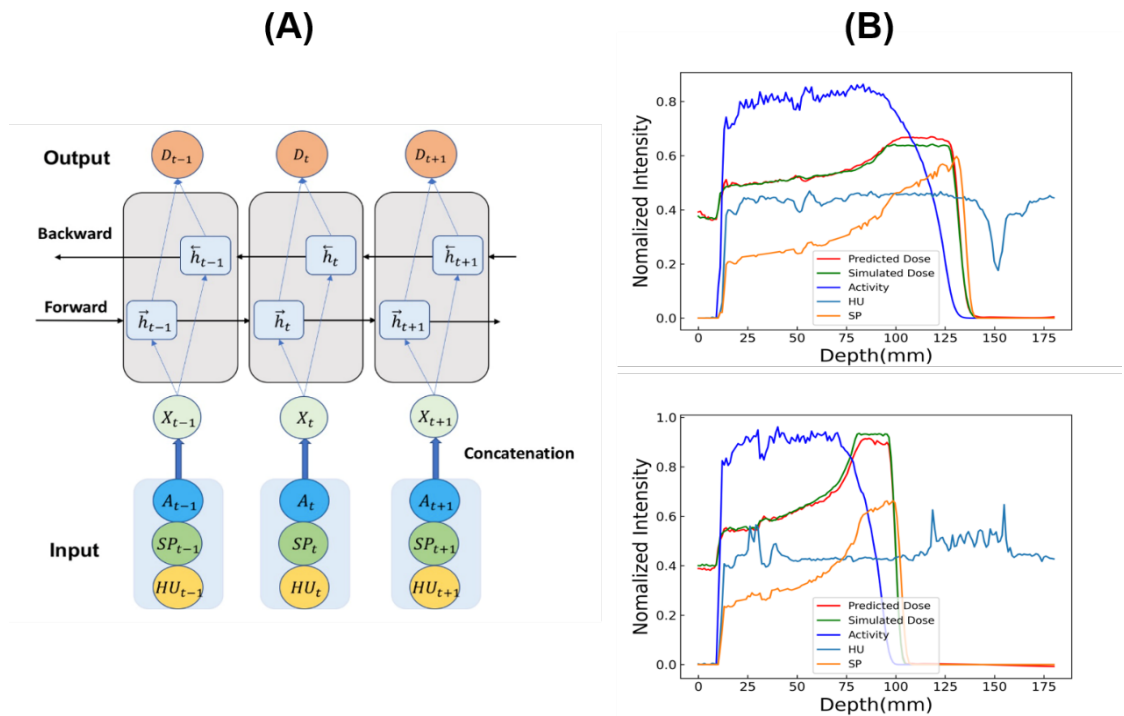

**Fig 2. (A)** The structure of a bi-directional RNN model. $X$ is an input vector by concatenating three values together (activity, HU and SP). $A_i$ is the activity profile at the $i_{th}$ depth. $h_i$ is the hidden unit of the RNN model at the $i_{th}$ depth. $D_i$ is the predicted dose at the $i_{th}$ depth. **(B)** Two examples of different SOBP widths and beam energies, each comprising three input profiles (activity, HU and SP) and two dose profiles (simulated and predicted) after normalization.

## 2.2 Machine Learning

Only 1D verification (central axis of a beamlet) was conducted in our study. The network architecture (bi-directional RNN model) and signal flow are shown in Fig. 2. HU information from CT image slices was used for tissue segmentation, density and elemental composition (C, H, O, etc.), based upon the stoichiometric conversion scheme in Table 1 (Schneider et al 2000). The ICRU tissue elemental compositions were used, and a conversion from atomic weight to mole fraction was conducted. Once the elemental composition of each voxel was determined, the stopping power (SP) (1D only) was analytically calculated using the CSDA model as described in the previous work (Hu et al 2020).

The complete data set (1819 cases) was divided into the training dataset (80%) and the testing dataset (20%). The total time steps of the RNN model was 250 and every model was trained for 150,000 epochs. All inputs and outputs (i.e. activity, HU, SP and dose profiles) were normalized respectively. The stochastic gradient descent (SGD) optimization was used. The initial learning rate was 0.005 and was multiplied by 0.95 every 5000 batches.

Three figures-of-merit were selected: $\Delta_{range}$, MRE and MAE. $\Delta_{range}$ represents the difference in the distal end of SOBP (80% dose fall-off) between predicted and simulated dose profiles. MAE (mean absolute error) reflects the absolute discrepancy between $\hat{y}_i$ (predicted dose) and $y_i$ (simulated/raw dose), while MRE (mean relative error) reflects the relative discrepancy. For each sample in the dataset, the maximum amplitude of its SOBP plateau, max (*SOBP*), was selected as the denominator in Eq. 1. To identify how the machine learning model performs at different depths, MAE and MRE were analyzed over two regions: a region including all the depths with >95% of max (*SOBP*) amplitude (referred to as MAE95 and MRE95), and a region including all the depths with >50% of max (*SOBP*) amplitude (referred to as MAE50 and MRE50).

$$\text{MAE} = \frac{1}{n}\sum_{i=1}^{n}|\hat{y}_i - y_i| \qquad \text{MRE} = \frac{1}{n}\sum_{i=1}^{n}\frac{|\hat{y}_i - y_i|}{\max{(SOBP)}} \qquad (1)$$

## 2.3 Biological Washout

The biological washout effect in a living body is important for range/dose verification. One avenue is to perform experiments under two conditions (live and dead animals), and the comparison between the two yields the information of time decay components. A brief summary of previous studies about washout is presented below, which includes two categories: 1) irradiation with radioactive beams, and 2) irradiation with standard beams.

In the first category (also known as the autoactivation technique), the radioactive beams were generated by the projectile fragmentation reaction through a target and selected by an isotope separator (Tomitani et al 2003, Mizuno et al 2003, Hirano et al 2013, Toramatsu et al 2018, Toramatsu et al 2020). For example, $^{12}$C beams can be first generated by a synchrotron and subsequently used to irradiate a $^7$Be target, producing two types of positron emitters $^{10}$C and $^{11}$C. For the autoactivation studies in tissues, the washout process was reported to comprise 2 or 3 components. For example, one study evaluated the metabolism of implanted $^{11}$C beams in rabbit thigh muscle and reported two decay components (Tomitani et al 2003). By examining the decay behaviors within the first 30 sec in $^{11}$C irradiation ($^{10}$C being produced as a by-product with a half-life of 1/60 of that of $^{11}$C), another study suggested three decay components (Mizuno et al 2003). For the brain irradiation, the fractions (half-lives) are $35 \pm 3\%$ ($2.0 \pm 1.8$ sec), $30 \pm 3\%$ ($140 \pm 18$ sec) and $35 \pm 1\%$ ($10191 \pm 2200$ sec). For the thigh muscle irradiation, the fractions (half-lives) are $30 \pm 4\%$ ($10.2 \pm 7.6$ sec), $19 \pm 3\%$ ($195 \pm 52$ sec) and $52 \pm 2\%$ ($3175 \pm 378$ sec), respectively. A plausible explanation was also provided (Mizuno et al 2003): the fast decay component is speculated to be associated with direct blood flow, the medium decay component may be associated with the penetration from tissue to blood vessels, while the slow decay component may be associated with the trapping of the nuclides by stable molecules of organs.

**Table1.** Parameters used for tissue segmentation and biological washout modeling. See more details about $k_1$ (for $^{15}$O) and $k_2$ (for $^{11}$C) in the Appendix section. Taking $^{15}$O for an example, $\lambda^1_{biological}$ (biological washout rate) is $0.06\ min^{-1}$ and $\lambda^1_{physical}$ (physical decay rate) is $0.34\ min^{-1}$, so that $k_1$ equals to $0.4\ min^{-1}$.

|  | $k_1$ | $k_2$ | C | O | HU |
|---|---|---|---|---|---|
| Air |  |  | 0 | 0.239 | [-1050,-950) |
| Lung |  |  | 0.105 | 0.749 | [-950,-120) |
| Adipose1 |  |  | 0.681 | 0.198 | [-120,-82) |
| Adipose2 | 0.34 | 0.034 | 0.567 | 0.308 | [-82,-52) |
| Adipose3 |  |  | 0.458 | 0.411 | [-52,-22) |
| Adipose4 |  |  | 0.356 | 0.509 | [-22,-8) |
| Adipose5 |  |  | 0.284 | 0.578 | [-8, 19) |
| Water | 0.4 | 0.039 | 0 | 0.882 | 0 |
| Tissue | 0.4 | 0.039 | 0.134 | 0.723 | [19,80) |
| Connective tissues | 0.34 | 0.034 | 0.207 | 0.622 | [80,120) |
| Bone1 |  |  | 0.455 | 0.355 | [120,200) |
| Bone2 | 0.37 | 0.039 | 0.423 | 0.363 | [200,300) |
| Bone3 |  |  | 0.391 | 0.372 | [300,400) |

In the second category, standard beam delivery is used and washout is modeled on the basis of pixels. It first segments voxels into different tissues based on HU values from CT images (Parodi et al 2007, Parodi et al 2008), and then assigns each tissue type different pre-known biological half-lives (one decay component). However, one limitation is that CT number alone is not sufficient to fully characterize chemical environment, a critical factor influencing the clearance of positron emitters (Knopf et al 2009, Ammar et al 2014). Along this path, an alternative is to leverage dynamic PET imaging to directly measure the biological washout rate per voxel (Espana et al 2010, Zhu et al 2011, Grogg et al 2015, Toramatsu et al 2020, Toramatsu et al 2022).

In our study, the second category was selected (see Appendix for more details) and only a single decay constant was modeled for each type of positron emitter ($^{11}$C or $^{15}$O), similar to one previous study (Grogg et al 2015). For instance, a decay rate of 0.03 min$^{-1}$ corresponds to a half-life time of 1386 s. Given the parameters in Table 1, the amplitude of the accumulated activity fluctuates about 5% over a period of 5 mins. Questions may be raised here regarding why several studies in the first category report a much higher degree of deviation in the measured activity between live (with washout) and dead (without washout) conditions (e.g. Fig. 6 in Hirano et al 2013, Fig. 4 in Toramatsu et al 2018). The difference stems from the modeling of a fast decay component (along with medial and slow components), as well as when acquisition starts. In our view, as long as the acquisition can start immediately after beam delivery and lasts 3-4 longer than the time constant of the fast decay component, the accumulated signal would not be very sensitive to whether one, two or three components are selected. Furthermore, a direct comparison is not meaningful here, since the biological washout will be quite different between two generating processes: target fragmentation reactions (clinical proton therapy) and irradiation with radioactive beams (Minzuno et al 2003).

Tissue segmentation was performed in the same manner as those in section 2.2. Based upon the published results of dynamic PET imaging (Grogg et al 2015), the parameters selected are summarized in Table 1. For $^{15}$O, the biological washout rate $\lambda^1_{biological}$ was set for three human tissues: soft tissue ($0.06 \, min^{-1}$), water ($0.06 \, min^{-1}$) and bone ($0.03 \, min^{-1}$) (zero for other tissue types due to the lack of data), and the physical decay rate $\lambda^1_{physical}$ was set to be $0.34 \, min^{-1}$. For $^{11}$C, the physical decay rate $\lambda^2_{physical}$ was set to be $0.034 \, min^{-1}$, and the biological decay rate $\lambda^2_{biological}$ was set to be $0.005 \, min^{-1}$ for three tissues (soft tissue, water and bone) and zero for others. We acknowledge that one limitation in our study is that the 1D model failed to incorporate spatial interference among voxels, that is, washout decreases activity at a given voxel but may increase activity at its neighboring voxels. This will be the topic to be investigated in the next step of our study.

In this section, only the scenario of activity only (no HU/SP included as input) was tested. A direct comparison was made between with and without washout (no depth selection applied). Since our study was intentionally devised to pinpoint the extent of performance degradation stemming from washout, the model was trained with the dataset applied with physical decay only and no model re-training was implemented.

**2.4 Depth Selection**

Sub-regions of different lengths (e.g. 100, 125 and 150 mm) were applied to the raw activity profiles (Fig. 3). For instance, a cutoff length of 100 mm means that only the activity profile (e.g. integral depth dose) between 0 and the 100 mm was kept. For each cutoff length, a unique RNN model was trained (no biological washout considered). Note that when a profile (e.g. low energy SOBP) has a range less than the cutoff length, it would not be affected by cutoff at all. However, when a profile (e.g. high energy SOBP) has a range beyond the cutoff length, data truncation and loss of information will make prediction more challenging. A heuristic for justifying our design is that given two different beam energies penetrating the same pathway inside a target, the activity profiles in a sub-region (e.g. proximal end) will be distinct from each other. By feeding the sub-region signals to the machine learning model, the mapping between activity and dose domains can be established, yet the accuracy is expected to deteriorate due to loss of information.

Like section 2.3, the scenario of activity only was tested (no HU/SP included as input). Neither physical decay nor biological washout was modeled in this section.

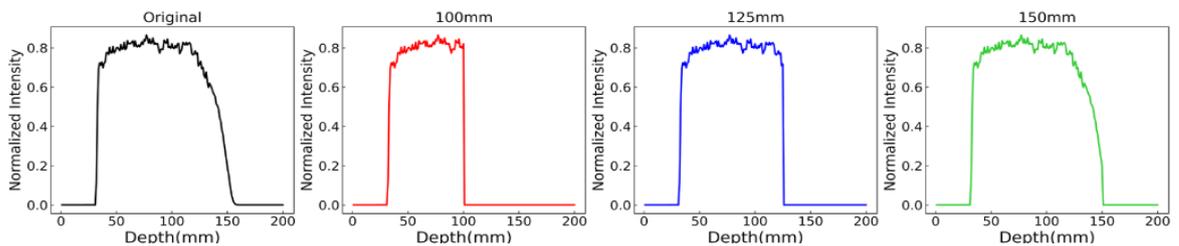

**Fig 3.** Illustration of the effect of depth selection on the activity profile for three cutoff lengths (100, 125, 150 mm).

**2.5 Worst Case**

The sensitivity analysis was conducted when both factors were present at the same time, referred to as the worst case. The cutoff length was set to be 125 mm. The biological washout was applied for a duration of 5 minutes. To highlight the impact of depth selection (data truncation) for higher beam energies, only a portion of the testing dataset (240 out of 360 samples having a range over 125 mm) were singled out for quantitative analysis. Two scenarios were tested (activity only, activty+HU+SP), and the RNN model of each scenario being trained separately. The inclusion of HU and SP as auxiliary features is a useful strategy to help improve the performance of machine learning, as already demonstrated in our previous studies (Hu et al 2020, Ma et al 2020).

# 3. Results

### 3.1 Biological Washout

Fig. 4A exemplifies the effect of biological washout, containing four profiles: activity (with and without washout), predicted dose and simulated dose. The difference between the activity profiles with and without washout is below 5.2% at a single depth. The falling edge is not influenced and overlaps well between each other. Note that the activity at a certain depth is the accumulated activity over a period of 5 mins.

Fig. 4B illustrates the spread of MRE95 and MAE95 of all samples in the testing dataset (360 samples), both increasing after the washout is introduced. As summarized in Table 2A, the mean MRE95 is 3.05% (without washout) and 4.50% (with washout), respectively. When extending to include more points on the proximal end, the mean MRE50 is 1.54% (without washout) and 2.59% (with washout), respectively. Despite the washout acting as a form of external perturbation, the RNN model is still able to function with accuracy degrading by a level of about 1-1.5%. In addition, the impact of biological washout on range prediction is even less significant, changing from 0.30 mm to 0.34 mm. It should be mentioned that although washout also affects spatial resolution, it is not expected to be significant relative to the intrinsic spatial resolution of a PET system (e.g. 4 mm in our study).

**Table 2.** Summary of quantitative results (based on 360 samples in the testing dataset) in terms of MRE95 and $\Delta_{range}$ (unit: mm): (A) 5 mins biological washout (no depth selection, the activity only scenario). (B) Different depth selections (no washout, the activity only scenario).

(A)

|  | MRE50 | MRE95 | MRE | $\Delta_{range}$ |
|---|---|---|---|---|
| **W/o washout** | 1.54% | 3.05% | 1.52% | 0.30 |
| **Washout (5 mins)** | 2.59% | 4.50% | 2.52% | 0.34 |

(B)

|  | MRE50 | MRE95 | MRE | $\Delta_{range}$ |
|---|---|---|---|---|
| **100 mm** | 2.50% | 5.85% | 2.40% | 2.47 |
| **125 mm** | 2.09% | 3.95% | 2.10% | 1.62 |
| **150 mm** | 1.55% | 3.12% | 1.53% | 0.53 |

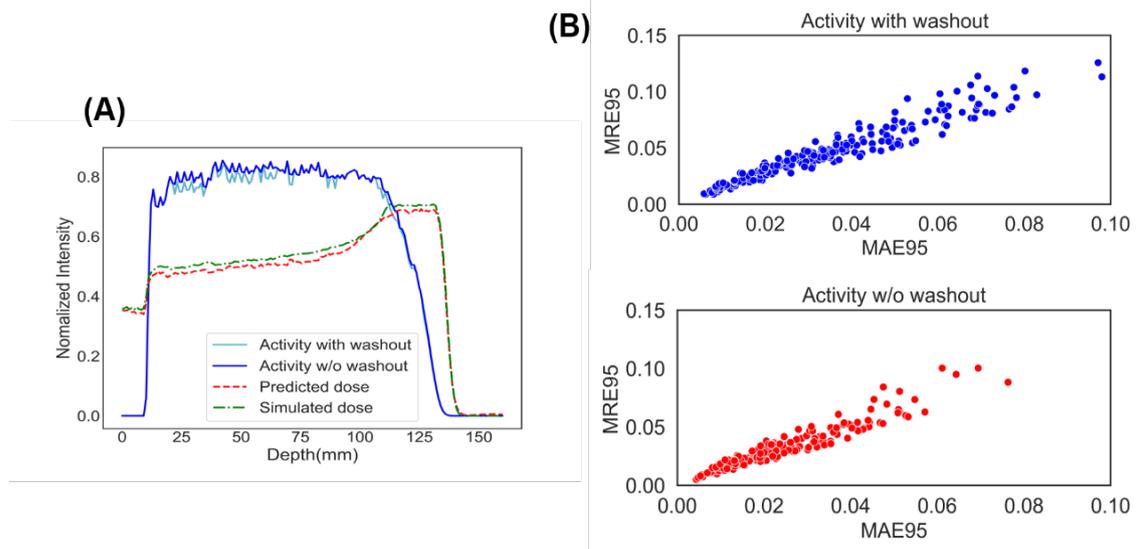

**Fig 4. (A)** Illustration of the effect of washout (the activity only scenario). **(B)** Comparison of MAE95 and MRE95 results between with and without biological washout (360 samples in the testing dataset).

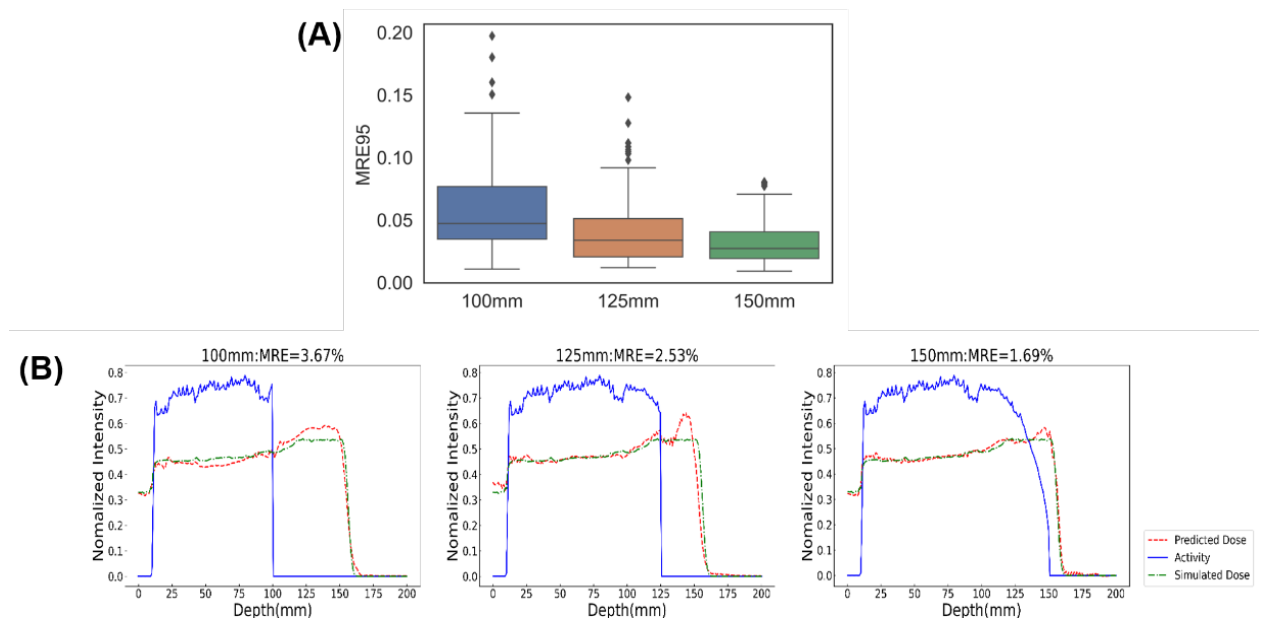

**Fig 5. (A)** MRE95 results of three cutoff lengths: 100 mm, 125 mm, and 150 mm (the activity only scenario, all 360 samples in the testing dataset). **(B)** Illustration of the performance with different cutoff lengths for one representative sample.

### 3.2 Depth Selection

The results are presented in Fig. 5 and Table 2B. The mean MRE95 is 5.85% (100 mm), 3.95% (125 mm) and 3.12% (150 mm), respectively. The mean $\Delta_{range}$ is 2.47 mm (100 mm), 1.62 mm (125 mm) and 0.53 mm (150 mm), respectively. Simply due to data truncation, the RNN model does demonstrate inferior performance. The implications are twofold. First, the testing dataset comprises samples of different ranges and the results reflect the gross impact of data truncation. Second, for those data samples (larger beam energy or range) more likely suffering from data truncation, the model functions despite inferior accuracy. This supports our initial hypothesis that by feeding the sub-region signals to the machine learning model, the mapping between activity and dose domains can still be established, yet the loss of information weakens the spatial correlation between the two domains.

**3.3 Worst Case**

To exemplify how biological washout and depth selection interleave with each other, the results of two samples are shown in Fig. 6. After the inclusion of additional features (acitivty+HU+SP), the model is found to be more immune to perturbation. First, the falling edges better align with each other between the simulated and predicted dose profiles. When only activity is used as input, the sample (top row) exhibits a shift of ~20 mm, but this is no longer observed in the acitivty+HU+SP scenario. Second, there exists noticeable oscillation over the SOBP plateaus for two samples when only activity is used as input. With the addition of HU and SP, the degree of oscillation decreases noticeably.

The distributions of quantitative results are shown in Fig. 7. For the activity only scenario, the mean $\Delta_{range}$ is 6.26 mm and the mean MRE95 is 10.1%. A number of samples even exhibit a range uncertainty of up to -30 mm. For the activity+HU+SP scenario, the mean $\Delta_{range}$ is 0.38 mm and the mean MRE95 is 3.11%. All samples exhibit a range uncertainty between -2 mm and 2 mm. This is not surprising from the perspective of machine learning. When more information (e.g. HU and SP) is available, the task of machine learning becomes easier and the RNN model is less sensitive to external perturbation.

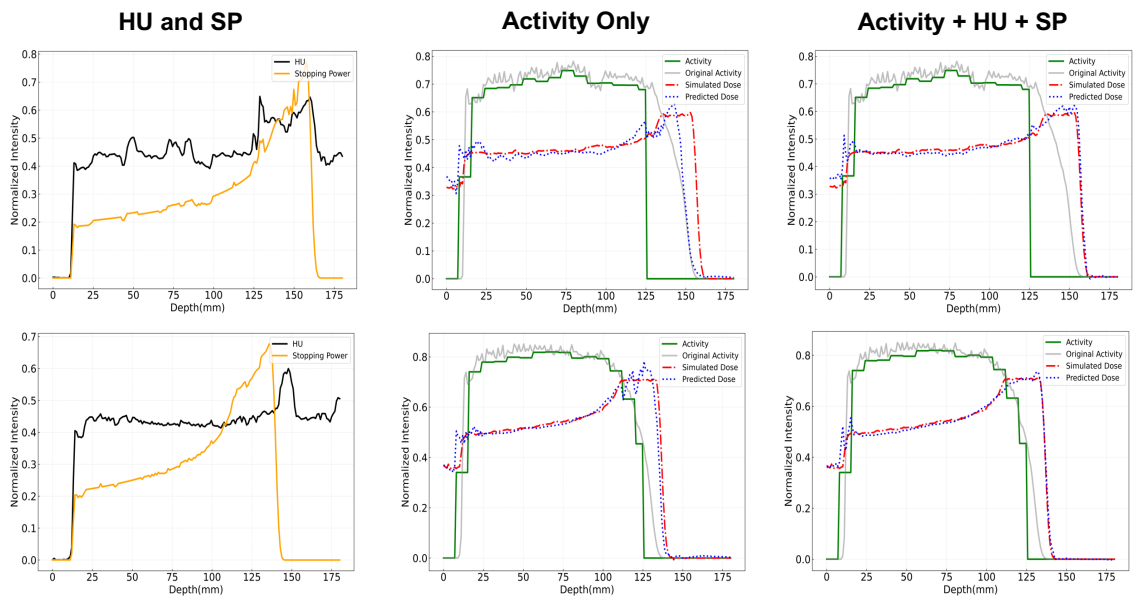

**Fig. 6.** The results of two representative samples in the worst-case (one sample per row). From left to right: normalized HU/SP, activity only, activity+HU+SP. The settings are 125 mm cutoff length and 5 mins washout. Note that washout was not incorporated in training the model.

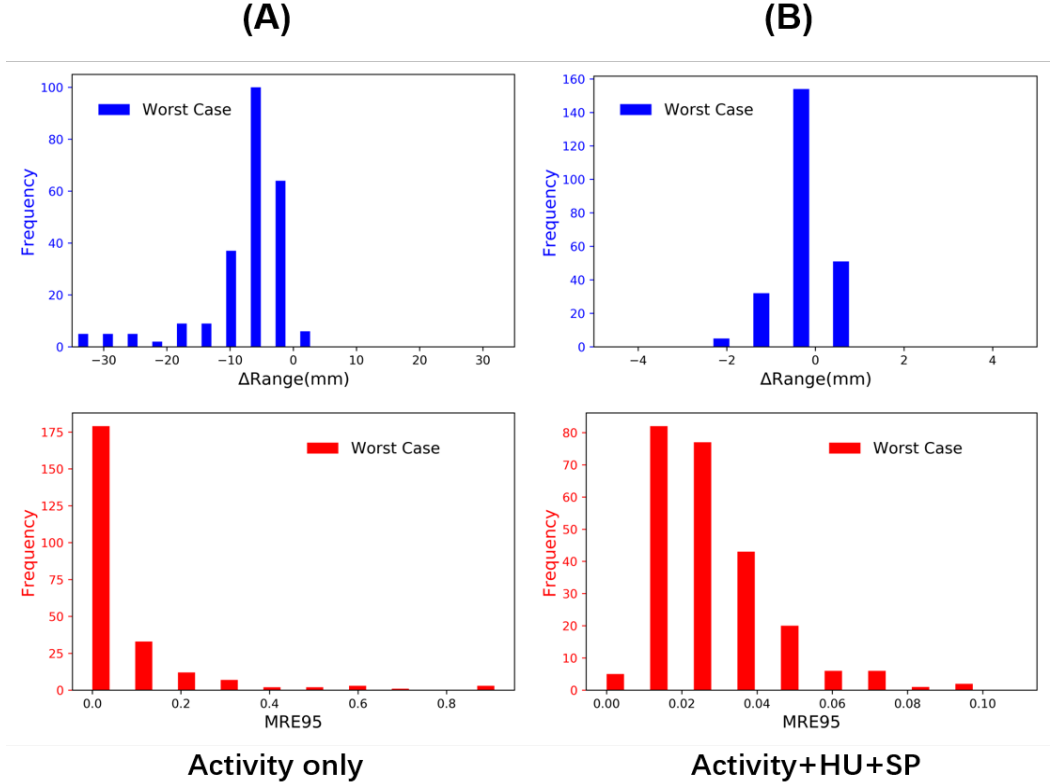

**Fig 7.** Distribution of MRE95 and $\Delta_{range}$ (removing the absolute value sign) for the worst-case scenario. **(A)** activity only **(B)** acitvity+HU+SP. Only a portion of the testing dataset (240 out of 360 samples with a range larger than 125 mm) were included for analysis.

As shown in Figs. 6 and 7, the incorporation of HU/SP further enhances accuracy in the worst case. When both factors are present at the same time, the correlation between input and output pairs becomes weakened. Whether a dose profile is predicted merely by SP prior rather than activity profiles (e.g. overfitting), is an intriguing question. To help answer that, a non-conventional validation task was added, by fixing HU and SP (based on one sample) and studying the performances of other samples. If the machine learning model makes a prediction only on SP prior, the predicted dose profiles would look roughly the same. This is not observed in our study as illustrated in Fig. 8. While sample 1 achieves good performance as expected (Fig. 8B), less accurate prediction is observed in Figs. 8C (sample 2) and 8D (sample 3). As the "faked" SP is more deviated from the "true" value, the degradation is much more noticeable (sample 2), one peak in the region of [100, 125] and another peak in the region of [125, 150]. When comparing Fig. 8B with Fig. 8D, sample 3 still obtains reasonably accurate prediction even with the "faked" SP, and the distal falling edges of dose profiles differ by almost up to

~20 mm relative to sample 1. This strongly suggests that in the acitvity+HU+SP scenario, the RNN model does efficiently utilize the important information embedded in activity profiles to make prediction, along with the assistance of HU/SP priors.

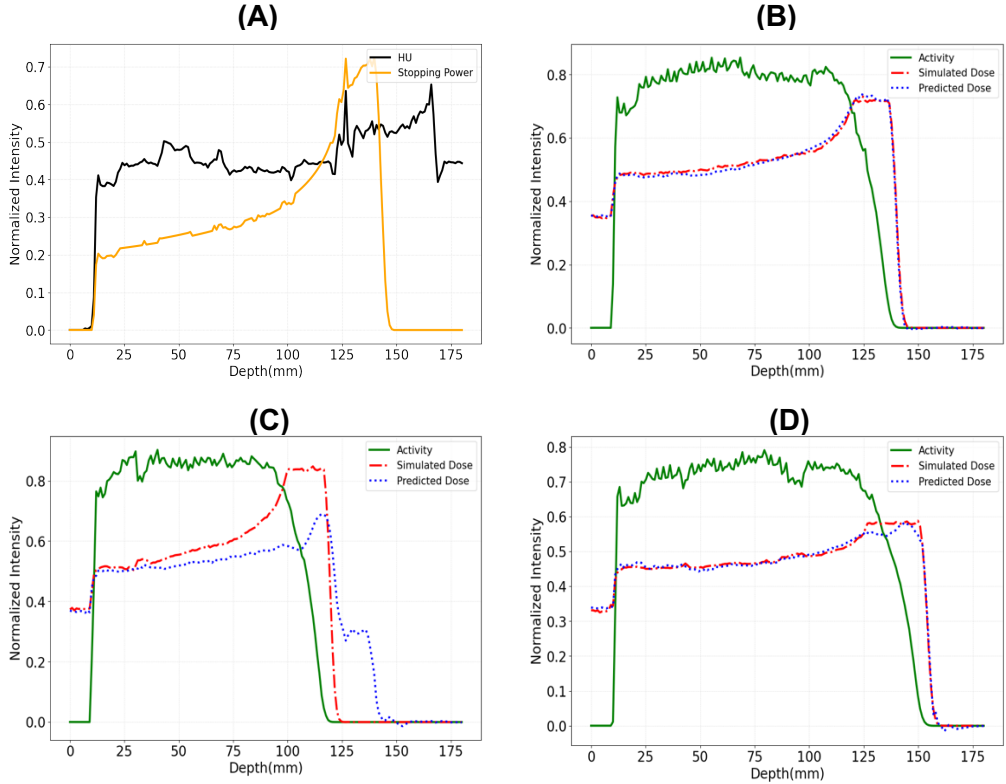

**Fig 8.** Validation results of three samples using constant HU and SP of sample 1. **(A)** Normalized HU and SP profiles (sample 1). **(B-D)** Performances of samples 1 to 3. For samples 2 and 3, their raw HU and SP profiles were intentionally replaced by the counterparts of sample 1.

## 4. Discussion

In this preliminary study, we conducted the sensitivity analysis of two factors (biological washout and depth selection), which potentially impact the performance of a machine learning framework for online range/dose verification. We believe the results will shed valuable light on a few aspects related to system prototyping and practical applications, such as spatial resolution, depth selection and data correction. For instance, low sensitivity to washout may simplify the workflow with respect to data processing and require no complicated data correction. Being able to maintain satisfactory performance after a moderate cutoff may render it possible to accomplish dose verification field by field in parallel. This manuscript is a logical extension of our previous studies and is the first step to integrate AI-based verification with external perturbations. Built upon the promising results of this study, we plan to tackle two additional tasks: 1) developing a 3D model of both SP prior and washout, 2) obtaining more reliable washout parameters in different tissue types through phantom and animal experiments.

**4.1 Biological Washout and Depth Selection**

The biological washout alters activity profiles, a process depending on the time elapsed after irradiation as well as tissue compositions. For the 1D model studied, its influence is found to be very limited in terms of MRE95 (less than 1.5%). As mentioned in section 2.3, three limitations require further validation in order to make the modeling more realistic. First, besides three selected tissues (bone, water and soft tissues) with available biological washout parameters in published literatures (Table 1), other tissue types were assumed to have zero washout. Second, the washout of $^{11}C$ in different tissues was assumed to be constant. Third, the washout was independently performed on the basis of each voxel, and interference among neighboring pixels was not yet incorporated. Nevertheless, given the maximum change introduced by washout is on the order of ~5% over a period of 5 mins, we envision that its influence may not be much different from noise added to the activity profiles. It is quite likely that the washout can be tolerated by the machine learning model, requiring no additional data correction other than physical decay and no model re-training. One previous study reported that washout effects introduce ~4 mm verification discrepancy between the measured and simulated range when the proton beam stops in soft tissue (Knopf et al 2009). It is thus desirable if our proposed machine learning framework is leveraged to improve the immunity to washout effect, through enhancing the correlation between activity with dose profile, particularly with the presence of both HU and SP as prior information.

The motivation of singling out depth selection as a factor, is to address the potential challenge of overlapped activities from multiple fields in proton therapy. Generating dataset and training a model for a single field is a much easier task compared to that for multiple fields, in terms of both difficulty and flexibility. The verification for the first field can be started while the second field is delivering beams. One promising finding is that after discarding activity beyond the cut-off length, the machine learning model still works despite slightly inferior accuracy. This confirms our initial assumption that the proximal activity distribution holds sufficient information for predicting the dose distribution on both proximal and distal ends. This may seem counter-intuitive because at those depths beyond the cutoff length, the activity profiles become zero. While awaiting further validation, we speculate that this may be due to the following two reasons: 1) beam energies not only determine distal fall-off positions (range), but also determine the production yield of positron emitters over the proximal region which are unaffected by cutoff. 2) we have re-trained different learning models for different cutoff lengths, otherwise the model will fail to function.

One accompanying issue is that only deep-seated targets (e.g. SOBP at >10 cm depth) were modeled in our study, which had extended entrance channels. Therefore, the conclusion drawn in this manuscript is not applicable to those treatment sites of shorter entrance channels (e.g. shallow-seated tumors). In other words, if the cutoff is too severe, the loss of information may be too significant to allow the machine model to predict, even with the assistance of HU and

SP features. Moreover, it is assumed in our study that the dose in the entrance channel delivered by a single field will not be affected by other fields. One might argue that this assumption does not hold in circumstances where a small angle is between two nearly parallel fields. If that happens, an immediate solution can be grouping together multiple fields/angles as a single field to generate dataset and perform verification.

**4.2 Practical Considerations**

Whatever secondary signals (i.e. positron emitters, prompt gamma, acoustic waves) to be used for verification in proton therapy, one common question arises as to what happens if the signals are systematically inaccurate. In this study, our intention is to investigate the influence of two factors on the raw activity (from MC simulation), which are both found to be tolerable. As outlined in the Introduction section, the ultimate accuracy of online verification relies on other aspects as well, such as cross section values of reaction channels and HU-based tissue composition derivation.

In this study, the machine learning model works with raw activity profiles from MC simulation (with a blurring kernel of 4 mm FWHM). In real situations, positron emitters will produce 511 keV gamma rays to be detected by radiation detectors as in standard PET imaging. The efficiency of the detection system, as well as scattering and random events of 511 keV gamma rays within patients, will impact the fidelity of activity profiles. This will comprise both signal-to-noise ratio (SNR) and spatial resolution. As shown in our previous study, the performance of machine learning indeed deteriorates when image SNR decreases and/or PET image reconstruction is introduced (Hu et al 2020). In other words, the activity profiles may not be as smooth as those in Fig. 4A in practical applications, depending on several factors such as total dose, detection efficiency, acquisition time and image reconstruction algorithm. Therefore, the sensitivity of machine learning to biological washout and depth selection may be aggravated relative to the findings in this manuscript. On a side note, proton FLASH is an emerging radiotherapy scheme that offers ultra-high dose rate delivery and ultra-fast irradiation (>40 Gy/sec), likely to be accompanied by hypo-fractionation (e.g. 5-10 Gy per fraction). This will be extremely beneficial for boosting the yield of proton-induced positron emitters (e.g. by a factor of ~2.5-5 relative to a standard fraction of 2 Gy), in case a higher SNR is desired for machine learning.

To implement our proposed approach, another three practical issues need to be considered. First, how the verification outcome (i.e. the comparison between the prescribed/simulated and predicted doses) can be utilized to alter a treatment plan and fine-tune spot weights, for the realization of adaptive proton therapy (Zhang et al 2022). Second, the signal acquisition time in our design (~5 minutes) is longer than typical treatment time (~1-2 minutes) and patient morphology may change during the course. One previous study reported that motion causes spatial deviation of up to 3 cm between measured and simulated activity distributions in abdominopelvic tumor cases (Knopf et al 2009). To what extent the machine learning model is sensitive to motion needs to be evaluated.

Third, a viable strategy needs to be devised for establishing a "gold standard", and we envision the following two possible approaches. On the one hand, dynamic PET imaging of proton-induced positron emitters in both phantoms and animals can be used. Parameters such as $^{11}$C and $^{15}$O production yield and biological washout rate, can be accurately validated voxel-by-voxel. On the other hand, to ensure simulated dose consistent with "true" dose, in-vivo dosimeters can be used at multiple points inside phantoms and animals. Both approaches will allow us to better check the fidelity of training dataset (inputs/outputs), and to fathom the performance limits of machine learning-based range/dose verification.

## 5. Conclusion

We investigated the influences of two factors (biological washout and depth selection) on a machine-learning based verification framework. The performance was quantitatively evaluated, in terms of $\Delta_{range}$, MRE and MAE. Overall, the framework shows good immunity to the perturbation associated with the two factors, maintaining good accuracy even in the worst-case. When more information (e.g. HU and SP) are available as auxiliary inputs, the task of machine learning becomes less sensitive to external perturbation. The detection of proton-induced positron emitters, combined with AI tools, has great potential to move towards online patient-specific verification in proton therapy.

**Appendix:**

We chose to use a simple analytical model of only one decay time constant, rather than 2 or 3 components as proposed in other studies. Assuming the acquisition can start immediately after beam delivery and lasts sufficiently long relative to the half-life times of positron emitters, the accumulated activity being used for machine learning would not be sensitive to the number of decay components (e.g. thinking about fitting a time-activity curve with one exponential function or the sum of two exponential functions).

The temporal behavior of the activity of positron emitters was modeled as below:

$$A(t) = \sum_i A_0^i \, e^{-\left(\lambda_{physical}^i + \lambda_{biological}^i\right)t} = \sum_i \lambda_{physical}^i N_0^i \, e^{-\left(\lambda_{physical}^i + \lambda_{biological}^i\right)t}$$

$$k_i = \left(\lambda_{physical}^i + \lambda_{biological}^i\right) \qquad (i=1 \text{ for } ^{15}\text{O}, i=2 \text{ for } ^{11}\text{C})$$

$$PET_{activity}(washout) = \int_0^T A(t)dt$$

$$= \sum_i N_0^i \left(1 - e^{-\left(\lambda_{physical}^i + \lambda_{biological}^i\right)T}\right) \frac{\lambda_{physical}^i}{\lambda_{physical}^i + \lambda_{biological}^i}$$

$$PET_{activity}(no\ washout) = \sum_i N_0^i \left(1 - e^{-\lambda_{physical}^i T}\right)$$

$$ratio = \frac{PET_{activity}(washout)}{PET_{activity}(no\ washout)}$$

$A(t)$ is the activity inside a single voxel. $A_0^i$ is the initial activity immediately after proton beam irradiation. The output of Monte Carlo simulation yields the number of positron emitters $N_0^i$ instead of activity $A_0^i$, although the choice of either $N_0^i$ or $A_0^i$ does not affect the results in our study because of normalization (i.e. only the ratio matters as shown in Fig. 4A). $\lambda_{physical}$ is the physical decay rate and $\lambda_{biological}$ is the biological washout rate. $PET_{activity}$ is the accumulated activity (after 5 mins) to be used as input for machine learning. The ratio (always less than one, tissue dependent) is what really matters as a form of external perturbation. The influence of washout is tied to the loss of activity. For the time being, Poisson noise (i.e. counting statistics) was not yet modeled.